\documentclass[preprint]{aastex}

\newcommand{\kms}{km\,s$^{-1}\,$}
\newcommand{\vlsr}{V$_{\rm LSR}$}

\newcommand{\laa}{{\lower0.8ex\hbox{$\buildrel <\over\sim$}}}
\newcommand{\gaa}{{\lower0.8ex\hbox{$\buildrel >\over\sim$}}}

\shorttitle{Hagiwara  et al.}
\shortauthors{Nuclear water maser emission in M\,51}

\begin{document}

\title{Water Maser Emission from the Active Nucleus in M\,51}

   \author{\sc Yoshiaki Hagiwara\altaffilmark{1}, 
           Christian Henkel\altaffilmark{1},
           Karl M. Menten\altaffilmark{1}, and
           Naomasa Nakai\altaffilmark{2}}

\altaffiltext{1}{Max Planck Institut f\"{u}r Radioastronomie,
                 Auf dem H\"{u}gel 69, 
                 D-53121 Bonn, Germany}
\altaffiltext{2}{Nobeyama Radio Observatory, 
                 Nobeyama  Minamimaki, Minamisaku, 
                 Nagano 384-1305, Japan}
%
\begin{abstract}
22\,GHz water vapor `kilomaser' emission is reported from the central 
region of the Whirlpool galaxy M\,51 (NGC\,5194). The red-shifted spectral 
features (\vlsr $\sim$ 560\,\kms), flaring during most of the year 2000, 
originate from a spatially unresolved maser spot of size $\laa$ 30\,mas 
($\laa$ 1.5\,pc), displaced by $\laa$ 250\,mas from the nucleus. The data 
provide the first direct evidence for the association of an H$_2$O kilomaser 
with an active galactic nucleus (AGN). In early 2001, blue-shifted 
maser emission (\vlsr $\sim$ 435\,\kms) was also detected. Red- and 
blue-shifted features bracket the systemic velocity asymmetrically. Within 
the standard model of a rotating Keplerian torus, this may either suggest 
the presence of a highly eccentric circumnuclear cloud or red- and 
blue-shifted `high velocity' emission from a radially extended torus. 
Most consistent with the measured H$_2$O position is, however, an association 
of the red-shifted H$_2$O emission with the northern part of the bipolar 
radio jet. In this scenario, the (weaker) northern jet is receding while
the blue-shifted H$_2$O emission is associated with the approaching southern 
jet. 

\end{abstract}



 
\section{Introduction}
 
The discovery of a thin molecular Keplerian torus surrounding a supermassive
object at the center of NGC\,4258 has proven that high-resolution imaging of 
water maser emission from active galaxies is a powerful method to investigate 
the dynamics of the circumnuclear region of AGN \cite{miy95, her99}. About 20 
water megamasers are known to date (e.g. Braatz, Wilson \& Henkel 1997), all of  which associate with Seyfert 2 or LINER nuclei. However, not all megamasers 
form part of a circumnuclear torus (e.g. Claussen et al. 1998; Gallimore et 
al. 2001; Hagiwara et al. 2001) and there are also galaxies that are characterized 
by much less luminous H$_2$O emission. Towards the inner 40$''$ of the starburst 
galaxy NGC\,253 and the nearby spiral M\,51 (NGC\,5194) weak ($L_{\rm  H_2O}$ 
$\sim$ 0.1 -- 1.0 L$_{\odot}$) water masers were observed, which were classified 
as `kilomasers' \cite{hop87, nak88}. Water masers with such intermediate 
luminosities are also observed in the Milky Way (e.g., W49N) and in the nearby 
galaxies IC\,10, M\,33, and M\,82 and are associated with HII regions \cite{gre90, 
arg94, bau96}. 

While all H$_2$O megamasers studied in detail were found to arise from the 
central few pc of active galaxies containing an AGN, it is an open question
whether or not some of the H$_2$O kilomasers can also be linked to the nuclear 
activity of galaxies. Are H$_2$O megamasers only the most luminous members 
of a much larger but less conspicuous population of nuclear masers as suggested 
by Ho et al. (1987)? The H$_2$O kilomaser source M\,51 is one of the nearest 
galaxies with an active Seyfert\,2/LINER nucleus \cite{ter98} and is thus a 
suitable target. To test the hypothesis of Ho et al. (1987) we present 
single-dish monitoring observations with the Effelsberg 100-m telescope and 
imaging observations with the VLA. 

\section{OBSERVATIONS}
 
\subsection{MPIfR 100-m radio telescope}

The Effelsberg 100-m telescope, operated by the Max-Planck-Institut f{\"u}r 
Radioastronomie, was used on several occasions between June 1995 and January 
2001. Between 1995 and 1997 observations of the 22\,GHz H$_2$O line were made 
in a position switching mode with offsets of 300$''$, employing a K-band maser 
receiver with $T_{\rm sys}$ $\sim$ 75\,K on an antenna temperature 
($T_{\rm A}^{*}$) scale. The backend was an autocorrelator covering a 
bandwidth of 25\,MHz with 1024 channels, yielding a velocity spacing of 
0.33\,\kms per channel. Since 2000 we used a dual channel K-band HEMT 
receiver with $T_{\rm sys}$ $\sim$ 60\,K (after averaging both channels) 
on a $T_{\rm A}^{*}$ scale. The observing mode was dual beam switching 
with a beam throw of 121$''$ and a switching frequency of $\sim$1\,Hz, 
providing baselines of high quality. The backend digital correlator 
consisted of eight modules (four for each orthogonally polarized channel), 
each of them providing 512 channels and a bandwidth of 40\,MHz (1.05\,\kms 
channel separation). Amplitude calibration was obtained by repeated 
measurements of 3C\,286 \cite{baa77, ott94}. The beam efficiency of the 
telescope was $\sim$0.33 at 22\,GHz. Pointing measurements were made 
towards the nearby continuum sources DA\,251 and 3C\,280; pointing 
errors were $<$15$''$. 

\subsection{Very Large Array}

The NRAO\footnote{The National Radio Astronomy Observatory (NRAO) is operated 
by Associated Universities, Inc., under a cooperative agreement with the 
National Science Foundation.}  Very Large Array (VLA) was used in its A 
configuration on January 23, 2001, to observe the 22 GHz H$_2$O line toward 
M\,51. The observations were made under good seeing conditions employing a 
single intermediate frequency band of width 12.5\,MHz divided into 64 channels 
of width 2.63\,\kms each. The band was centered at \vlsr = 560\,\kms, which 
is $\sim$90\,\kms\ redshifted to the systemic velocity of the galaxy. 
Considering filter roll-off at the edges of the passband, the usable velocity 
range was 495 -- 625\,\kms. Given that at the time of our VLA observations 
H$_2$O emission at velocities lower than the systemic velocity had not yet 
been detected and considering the constraints imposed by the VLA's spectral 
line correlator modes, our setup did not cover the \vlsr-range of that 
low-velocity H$_2$O emission. During the local sidereal time interval 
$12\rlap{.}^{\rm h}5$ to $15^{\rm h}$ the nuclear position of M\,51 was 
observed in a sequence of 5 min. scans, interspersed with 2 min. scans 
of the calibrator source J1419+543 for which we used a position of 
$(\alpha,\delta)_{\rm J2000}$ = $14^{\rm h}19^{\rm m}46\rlap{.}^{\rm s}5974$, 
$+54^\circ23'14\rlap{.}{''}787$, determined with sub-milliarcsecond 
precision by Johnston et al. (1995). The resultant integration time on 
M\,51 is about 100 min. The flux density, $S$, scale was determined by 
observations of 3C\,286 and is estimated to be accurate to within 10\%, 
assuming that  $S({\rm 3C\,286}) =$ 2.54\,Jy at 22.2\,GHz. The data were 
calibrated and mapped in the standard way using the NRAO's AIPS. The 
synthesized beam has a FWHM size of $0\rlap{.}{''}102\times0\rlap{.}{''}096$, 
corresponding to a linear scale of $\sim$5\,pc at $D$ $\sim$10\,Mpc, and 
a position angle of $-$7$\rlap{.}{^\circ}6$. Maps of the H$_2$O emission 
were made for each velocity channel. These `channel maps' have a typical 
$1\sigma$ rms noise level of about 4\,mJy\,beam$^{-1}$. We also produced 
a map of the H$_2$O emission integrated over the velocity range over which 
significant emission was found in the channel maps, i.e. between 538 and 
590\,\kms. This map has an rms noise level of 2.8\,mJy\,beam$^{-1}$. 

\section{Results}

Fig.\,\ref{fig1} shows the spectra obtained between 1995 and 2001. The upper
panel displays single dish profiles showing particularly strong maser emission
in spring 1997 and 2000. The latter flare lasted until almost the end of the 
year. Most of the emission is seen red-shifted relative to the systemic 
velocity ($V_{\rm sys,LSR}$ = 469$\pm$5\,\kms; Kuno et al. 1995). For 
the first time blue-shifted emission is also observed: Clearly seen at 
the end of January 2001, the strongest component at $V_{\rm LSR}$ = 
430.5\,\kms\ is also visible in the Dec. 2000 spectrum. The blue-shifted 
emission covers a range of 410 to 440\,\kms\ or --60 to --30\,\kms\ relative 
to $V_{\rm sys}$, while the red-shifted emission is observed between 538 and 
592\,\kms\ or about +70 to +120\,\kms\ relative to $V_{\rm sys}$. Integrated 
intensities determined for the velocity intervals 410 -- 440\,\kms\ and 
538 -- 592\,\kms\ are not correlated. Although not shown, all seven 
epochs include the velocity range near 690\,\kms, where Nakai et al. 
(1995) detected emission in early 1995. No H$_2$O feature is seen at their
velocity in our spectra.

The lower panel shows the VLA spectrum taken on Jan 23, 2001. These are the 
first molecular data from M\,51 taken with subarcsecond resolution. Both 
with respect to flux density and lineshape, the spectrum is consistent with 
the single-dish data from December 2000 and January 2001. The 540, 566, and 
578\,\kms\ components are all seen. Within the limits of noise, there is no 
missing flux. The emission arises from a spatially unresolved maser spot of 
size $\laa$ 100\,mas at a position of $(\alpha,\delta)_{\rm J2000}$ = $13^{\rm 
h}29^{\rm m}52\rlap{.}^{\rm s}7085$ $\pm$ $0\rlap{.}^{\rm s}0004$, 
$+47^\circ11'42\rlap{.}{''}796$ $\pm$ $0\rlap{.}{''}013$. The quoted 
uncertainty is the quadratic sum of the error due to the thermal noise in 
the map and the upper bound of the systematic error caused by the fact that 
the calibrator source and M51 are at different positions in the sky. This 
upper bound, which dominates the error budget, was determined by producing a 
map of 3C286 by applying calibration solutions determined from J1419+543 scans 
and determining the offset between 3C286's centroid position in that map and 
its true position. Since the angular separation between J1419+543 and 3C286 
is significantly greater than the separation between J1419+543 and M51, and 
since M51 was observed over a much wider hour angle range than 3C286, we 
consider the above offset as a conservative upper bound on the position error 
of the M51 H$_2$O emission. The absolute position errors of J1419+543 and 
3C286 are negligible (see e.g. Johnston et al. 1995; Ma et al. 1998). 

In none of the channel maps (see Sect.\,2.2) does the centroid position of 
the H$_2$O emission significantly deviate from the position determined for 
the integrated emission. From the measured peak flux density and the beam 
size we obtain a lower limit to the maser's brightness temperature of 10$^{4}$\,K. Assuming Gaussian distributions for both the H$_2$O emission and the beam, 
deconvolution yields a maser source size of $\laa$ 30\,mas and a peak brightness temperature of $\gaa$ 10$^{5}$\,K. 

To search for 22.2 GHz continuum emission, we made a map from a 
$u$,$v$-database obtained by averaging the line-free channels. No continuum 
emission was detected at a $1\sigma$ noise level of 0.52\,mJy\,beam$^{-1}$.


\section{Discussion}

\subsection{H$_2$O maser versus radio continuum peak position}
 
The most important result of our observations is the determination of an 
absolute position of the maser with a 1$\sigma$ accuracy of $\sim$15\,mas. 
A comparison of this maser position with an 8.4\,GHz VLA A-array 
continuum peak position 
($(\alpha,\delta)_{\rm J2000}$ = $13^{\rm h}29^{\rm m}52\rlap{.}^{\rm s}7101$ 
$\pm$ $0\rlap{.}^{\rm s}0016$, 
$+47^\circ11'42\rlap{.}{''}696$ $\pm$ $0\rlap{.}{''}026$; Kaiser et al. 2001)
shows that the water kilomaser arises from a hotspot $\sim$100\,mas 
(5\,pc at $D$ = 10\,Mpc) north of the radio continuum peak that marks the 
nuclear position of the galaxy (e.g. Grillmair et al. 1997, 1998). A comparison 
with a 5\,GHz VLA A-array continuum map of slightly lower angular resolution
by Crane \& van der Hulst (1992) also shows good agreement; in this case 
the position of the continuum peak is $(\alpha,\delta)_{\rm J2000}$ = 
$13^{\rm h}29^{\rm m}52\rlap{.}^{\rm s}706$, $+47^\circ11'42\rlap{.}{''}55$  
(no errors given), i.e. the maser is located $\sim$25\,mas east and 
$\sim$250\,mas north of the compact continuum source. From even lower
resolution 5\,GHz VLA B-array and 15\,GHz C-array data, Turner \& Ho (1994)
determined a continuum peak position of $(\alpha,\delta)_{\rm J2000}$ = 
$13^{\rm h}29^{\rm m}52\rlap{.}^{\rm s}711$, $+47^\circ11'42\rlap{.}{''}61$;
an absolute positional uncertainty of 100 -- 500\,mas is quoted. A comparison 
with our data suggests that the maser is arising from a position 25\,mas to 
the west and 185\,mas to the north of the continuum peak. Since the registration 
of one map to another is problematical when different calibrators and frequencies 
are involved, we should consider the quoted individual position errors with 
utmost caution. A better approach is to compare the three offsets derived: 
With an average offset of 180$\pm$80\,mas, this is first direct evidence for a 
{\it kilomaser} located {\it in the vicinity of an AGN}. 

\subsection{Extended molecular complex, nuclear torus or nuclear jet?}

The nuclear region of M\,51 is complex. CO $J$=1--0 and 2--1 interferometric 
data (Scoville et al. 1998; Aalto et al. 1999) show a $\sim$2$''$ sized 
centrally located molecular cloud with a bulk velocity that is red-shifted 
with respect to $V_{\rm sys}$. Is this peculiar cloud responsible for the 
predominance of redshifted H$_2$O maser features? This is not likely. The 
average velocities of the main CO sub-complexes are in the range 462\,\kms 
$<$ $V_{\rm LSR}$ $<$ 545\,\kms\ \cite{sco98} and do not cover the velocity 
range seen in H$_2$O (Fig.\,\ref{fig1}). At $V_{\rm LSR}$ $\sim$ 565\,\kms, 
the characteristical H$_2$O velocity of the red-shifted features, Scoville 
et al. (1998) find CO $J$=2--1 emission 1$''$ south-west of the nucleus, well 
outside the positional error bars of our data (Sect.\,4.1). According
to Scoville et al. (1998; their Sect.\,3) the asymmetry in the CO distribution 
is significant, with the bulk of the emission being red-shifted with respect 
to the systemic velocity and arising from western offsets with respect to the 
nuclear position. 

HCN $J$=1--0 line emission is weaker but traces gas of higher density
and may thus be a more appropriate tracer of the warm dense molecular gas
that gives rise to 22\,GHz H$_2$O emission. In the HCN $J$=1--0 line, an
asymmetry favoring red-shifted over blue-shifted emission (as seen in CO) 
is not apparent. At the 3\,$\sigma$ level, Kohno et al. (1996) find HCN 
$J$=1--0 emission at 551--582\,\kms\ in the inner 5$''$ of M\,51 and identify 
a dense nuclear torus with $R$ $\sim$70\,pc and an apparent $V_{\rm rot}$ = 
16\,\kms\ at its outer edge; the estimated inclination is $i$ $\sim$ 50$^{\circ}$ 
-- 80$^{\circ}$ (much higher than the inclination of the large scale 
disk). Red-shifted emission with respect to $V_{\rm sys}$ is seen to the 
west, blue-shifted emission arises east of the nucleus. With an elongation 
along P.A. $\sim$ 70$^{\circ}$, the rotation axis (P.A. $\sim$ 160$^{\circ}$) 
is aligned with the nuclear jet. If the nuclear torus shows Keplerian rotation 
and if the very center dominates the mass ($\sim$10$^{7}$\,M$_{\odot}$), 
$V_{\rm rot}$ = $V$ -- $V_{\rm sys}$ $\sim$ 95\,\kms\ (for the H$_2$O line 
velocities, see Sect.\,3) should be reached at $R$ $\sim$ 2\,pc (40\,mas). 
In view of the positional error bars (Sect.\,4.1) this is not inconsistent with 
the observational data, but we note that H$_2$O emission from $V$ -- $V_{\rm sys}$ 
$\sim$ 95\,\kms\ would then arise from a location 40\,mas west, not (as our 
data more likely suggest) $\sim$180\,mas north of the nucleus. Assuming that 
the blue- and red-shifted masers originate from those parts of the torus that 
are viewed tangentially, the asymmetry in $V$ -- $V_{\rm sys}$ (Sect.\,3) infers 
a dense warm torus that is radially extended or the existence of highly 
eccentric orbits. The position offset between red- and blue-shifted features 
should be detectable with the A-array of the VLA.

Instead of inferring that H$_2$O is tracing a nuclear torus, we may 
postulate that the emission is associated with the radio jet. 
As already mentioned, the radio jet and the rotational axis of the inner
(HCN) torus are characterized by P.A. $\sim$ 160 -- 165$^{\rm \circ}$. 
The maser position determined in Sect.\,4 is not far from this axis 
(for the line connecting the 8.4\,GHz continuum peak (Kaiser et al. 2001)
with the maser position, P.A. = 174$^{\circ}$ $\pm$ 10$^{\circ}$). 
Agreement with measured positions is thus better than in the Keplerian 
disk scenario (for the 8.4\,GHz peak, the distance to the jet axis is 
15$\pm$35\,mas). In view of unifying schemes predicting a Doppler boosted 
approaching and a Doppler dimmed receding jet, it is likely that the 
southern jet is approaching, while the northern jet is receding. 
Within this context the red-shift of the H$_2$O emission is readily 
explained: Entrainment of the molecular gas by the jet may cause the 
velocity offset from systemic.

It is well known that 22\,GHz H$_2$O emission is often associated with shocks
(e.g. Elitzur 1995) that may raise not only kinetic temperatures but also 
H$_2$O abundances. Such shocks can be triggered (on small scales) by jets 
arising from young stellar objects or (on larger scales) by nuclear radio 
jets. Megamasers not oriented perpendicular but along the nuclear jets 
have already been observed towards NGC\,1052 \cite{cla98}, Mrk\,348 
(NGC\,262; Peck et al. 2001), NGC\,1068 (the `jet masers'; Gallimore et 
al. 2001), and Circinus (Greenhill et al. 2001) and may either arise from a 
fortuitously placed foreground cloud or, more likely, from a shocked dense 
region near the boundary of the jet. According to this latter scenario, 
the blue-shifted features from M\,51 (see Fig.\,1) should be associated 
with the southern jet. This can be checked observationally, since position 
offsets should be large enough to be detected by VLA A-array observations.

\subsection{Conclusions}

Presenting the first direct evidence for a nuclear H$_2$O kilomaser, our 
data are not inconsistent with the suggestion of Ho et al. (1987) that 
there must be a family of weak nuclear `megamasers'. {\it The kilomaser 
in M\,51 shows characteristic properties of megamasers}. Only the measured 
luminosity is smaller. Interpreting slight offsets between nuclear and 
maser position and postulating that the northern jet is receding, we conclude 
that the masers in M\,51 are most likely associated with the nuclear jet. 
Positional uncertainties are, however, too large to definitely rule out 
emission from a circumnuclear disk.

\acknowledgments

We wish to thank W. A. Baan for providing a high precision J\,2000.0 
8.4\,GHz continuum peak position of M\,51, W. A. Altenhoff for
the selection of a sufficiently accurate program to convert coordinates 
from B\,1950.0 to J\,2000.0, and A. B. Peck for critically reading the 
manuscript. We also appreciate  P. T. P. Ho for his suggestions on the manuscript.

\clearpage


%
%
\figcaption[m51fig1.ps]{Upper panel: 22\,GHz H$_2$O spectra taken with 
the Effelsberg 100-m telescope between June 1995 and January 2001 from 
M\,51. Note the enlargement of the temperature scale by factors of 2 
or 3 for the most recent spectra marking the end of the flare.
For the first two epochs channel spacings are 1.3\,\kms; for the other
spectra the spacing is 1.05\,\kms. 1$\sigma$ noise levels are, in 
chronological order, 22, 17, 28, 15, 15, 13, and 9\,mJy. Lower panel: 
VLA spectrum from January 23. For details, see Sect.\,2.2. \label{fig1}}
%
%
%
%
%




\clearpage

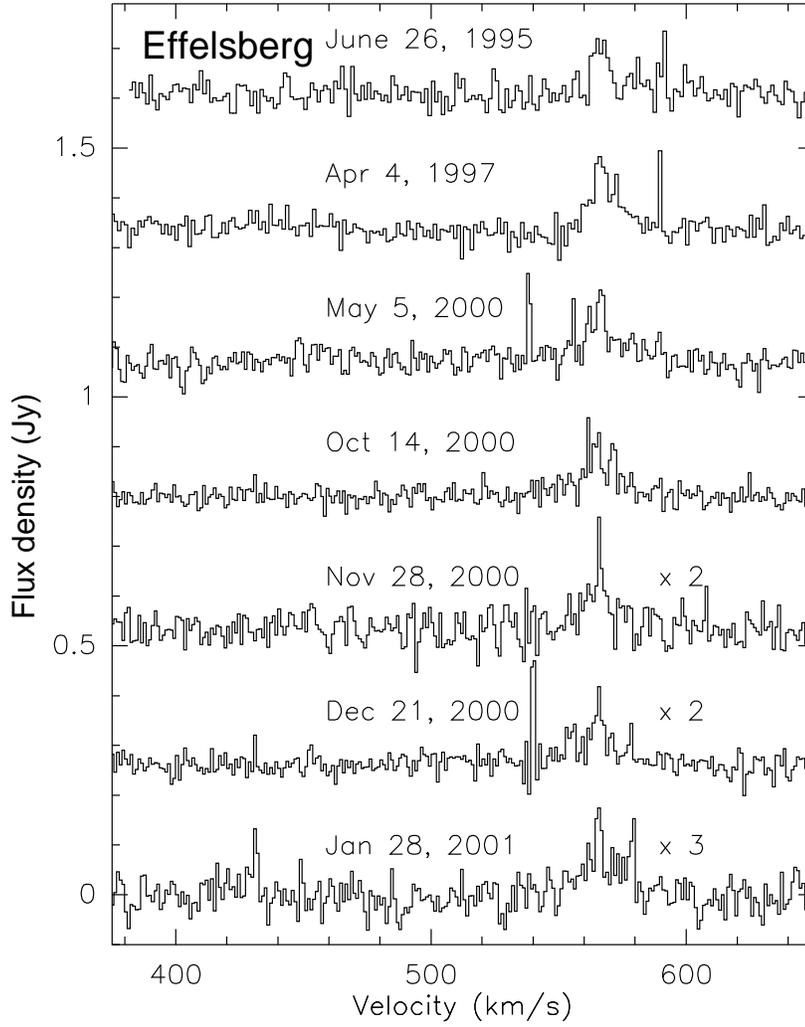
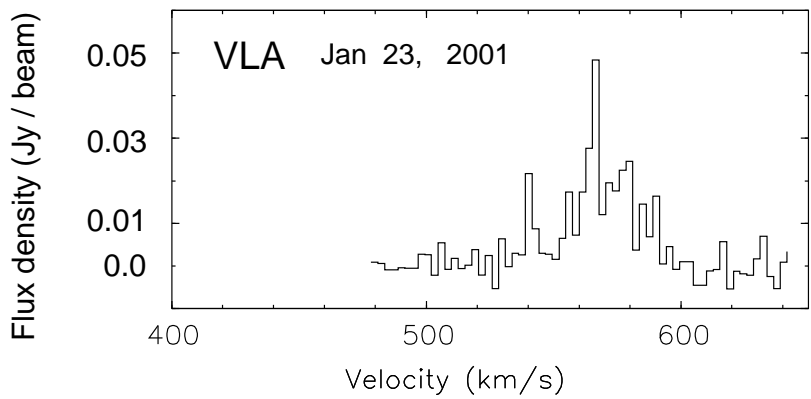
\end{document}